\documentclass[preprint]{aastex61}

\def\arcsec{\hbox{$^{\prime\prime}$}}

\submitjournal{Astrophysical Journal Letter}
\received{2017.4.16}
\accepted{2017.5.2}

\shorttitle{Plasmoid ejection observed with ALMA}
\shortauthors{Shimojo et al.}

\begin{document}

\title{FIRST ALMA OBSERVATION OF A SOLAR PLASMOID EJECTION FROM AN X-RAY BRIGHT POINT}

\correspondingauthor{Masumi Shimojo}
\email{masumi.shimojo@nao.ac.jp}

\author[0000-0002-2350-3749]{Masumi Shimojo}
\affil{National Astronomical Observatory of Japan, Tokyo, 181-8588, Japan}
\affil{Department of Astronomical Science, The Graduate University for Advanced Studies (SOKENDAI), Tokyo, 181-8588, Japan}

\author{Hugh S. Hudson}
\affil{School of Physics and Astronomy, University of Glasgow, Glasgow, G12 8QQ, Scotland, UK}
\affil{Space Sciences Laboratory, University of California, Berkeley, CA 94720, USA}

\author{Stephen M. White}
\affil{Space Vehicles Directorate, Air Force Research Laboratory, Kirtland AFB, NM 87117-5776, USA}

\author{Timothy S. Bastian}
\affil{National Radio Astronomy Observatory, Charlottesville, VA 22903, USA}

\author{Kazumasa Iwai}
\affil{Institute for Space-Earth Environmental Research, Nagoya University, Nagoya, 464-8601, Japan}
\affil{National Institute of Information and Communications Technology, Tokyo, 184-8795, Japan}

\begin{abstract}

Eruptive phenomena such as plasmoid ejections or jets are an important feature of solar activity with the potential for improving our understanding of the dynamics of the solar atmosphere. Such ejections are often thought to be signatures of the outflows expected in regions of fast magnetic reconnection. The 304 \AA \ EUV line of Helium, formed at around 10$^{5}$ K, is found to be a reliable tracer of such phenomena, but the determination of physical parameters from such observations is not straightforward. We have observed a plasmoid ejection from an X-ray bright point simultaneously at millimeter wavelengths with ALMA, at EUV wavelengths with AIA, in soft X-rays with Hinode/XRT. This paper reports the physical parameters of the plasmoid obtained by combining the radio, EUV and X-ray data. As a result, we conclude that the plasmoid can consist either of (approximately) isothermal $\sim$10$^{5}$ K plasma that is optically thin at 100 GHz, or else a $\sim$10$^{4}$ K core with a hot envelope. The analysis demonstrates the value of the additional temperature and density constraints that ALMA provides, and future science observations with ALMA will be able to match the spatial resolution of space-borne and other high-resolution telescopes.

\end{abstract}


\section{Introduction} \label{sec:intro}

Eruptions are one of the key mechanisms of explosive energy release in the solar atmosphere.  Since magnetic reconnection models predict that ejections of magnetic islands from current sheets accompany fast magnetic reconnection \citep[e.g.,][]{Priest85,Shibata01,Bhat09}, plasmoid ejection from flare sites have been investigated intensively \citep[e.g.,][]{Shibata95,Ohyama98,Hudson01,Takasao16}. Observationally, an ejected plasmoid is identified as a moving bright blob, sometimes embedded in another plasma flow such as a coronal jet \citep{Alex99}. Plasmoid ejections can also occur in events much smaller than flares, such as polar jets \citep{Shimojo07,Moore10,Ster15} and transition-region explosive events \citep{Innes09,Innes15}. Images in the 304 \AA \ line of He II have proven to be a convenient way to study ejecta in small events, especially in polar coronal jets, thanks to the high spatial resolution of the Atmospheric Imaging Assembly \citep[AIA:][]{AIA12} telescope on the Solar Dynamics Observatory \citep[SDO:][]{SDO12} satellite. This is a transition region line, implying that the ejecta have temperatures of order 10$^{5}$ K. Understanding the energetics of an ejection and the relationship to other energy sinks and sources in such an event is important for understanding energy release processes. However, it is difficult to use the He II line data for quantitative analysis because the intensity of the line depends strongly on resonance scattering of ambient EUV and X-ray radiation \cite[e.g.,][]{Labr12}.

New opportunities for observing phenomena in the Sun's atmosphere are being provided by the Atacama Large Millimeter/sub-millimeter Array \citep[ALMA:][]{ALMA10}. Emission from the Sun at millimeter wavelengths consists of non-flaring thermal emission produced under local thermal equilibrium (LTE) conditions, as well as synchrotron emission from electrons accelerated to MeV energies in solar flares \citep[e.g.,][]{White92}. Millimeter observations of the quiet Sun are optically thick in the solar chromosphere, and provide direct temperature measurements that are straightforward to interpret. 

An ALMA observation of a solar active region at 100 GHz on 2015 December 17 has been found to include a plasma ejection from an X-ray bright point (XBP). The event (SOL2015-12-17T19:44) was also observed by AIA and the soft X-Ray Telescope \citep[XRT:][]{XRT07,XRTCam08} on the Hinode satellite \citep{Hinode07}. In this paper, we analyze this first ALMA observation of such an ejection, and examine its thermal structure from the combination of 100 GHz, EUV and X-ray data. In section 2, we give the details of the observations with ALMA, AIA and XRT. In section 3, we describe the plasmoid ejection observed with these telescopes. We then discuss three basic working hypotheses to explain the 100 GHz, EUV and X-ray emissions from the plasmoid, and test them based on the observational data. Finally, we discuss the thermal structure of the plasmoid.

\section{Observation} \label{sec:obs}

The event occurred near the large leading sunspot of active region NOAA12470, during one of the test observations for an ALMA solar commissioning campaign. The dataset has been provided by the ALMA observatory as a Science Verification (SV) data release. The central frequency of the solar observations is 100 GHz. We calibrated the visibility data following the method described in \citet{Shimojo17}, and synthesized images every 2 seconds. The telescope array included 22 $\times$ 12m antennas and 9 $\times$ 7m antennas during this observation. The major and minor axes of the resulting synthesized beam are 6.3\arcsec \ and 2.3\arcsec, respectively. Small-scale jitter caused mainly by fluctuations in the Earth's atmosphere is evident in the 2-second cadence movie made from the ALMA data after the standard calibration procedure. To eliminate this motion, we carried out self-calibration using the image synthesized from the entire observing period as a reference, and this successfully removed the artificial motion. The noise level of the synthesized 2-second 100 GHz images is estimated using the method in \citet{Shimojo17}, and found to be $\sim$11 K.

\begin{figure}
\plotone{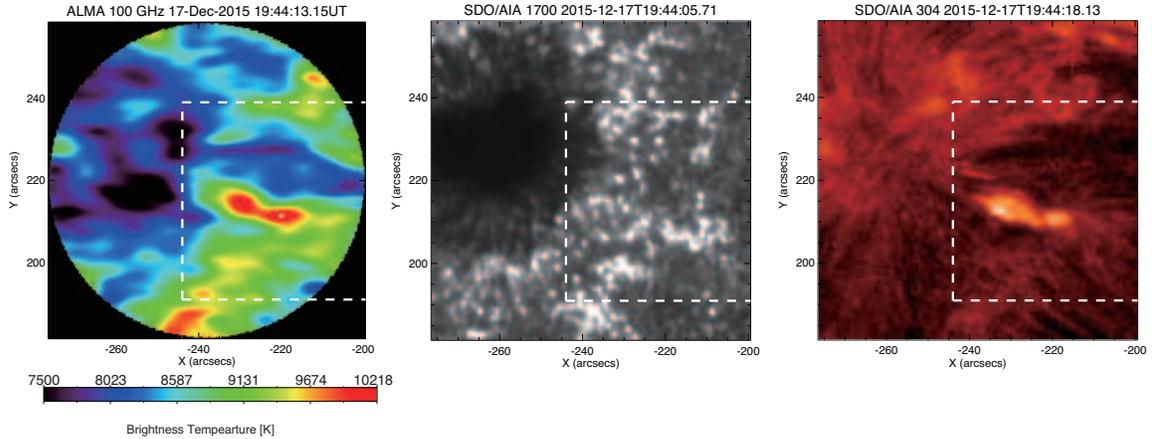}
\caption{The leading sunspot of active region NOAA12470 observed with ALMA at Band 3 (100 GHz, left panel), in ultraviolet continuum from the lower chromosphere (AIA 1700 \AA, middle panel), and the He II transition region line (AIA 304 \AA, right panel). The dashed squares in the panels indicate the area displayed in Figure \ref{fig:fig2} and Movie \ref{mov:mov1}.\label{fig:fig1}}
\end{figure}

Since the field of view of ALMA ($\sim$60\arcsec \ at 100 GHz) is significantly smaller than the Sun, the interferometer images do not measure the background level of quiet-Sun emission. Single-dish observations of the full Sun were therefore taken to provide the absolute temperature scale \citep{White17}. The temperature scale in the left panel of Figure \ref{fig:fig1} shows the result of correcting the interferometer data for the background temperature measured at the observing location in the concurrent single dish data (whose resolution is the same as the 100 GHz field of view, $\sim$60\arcsec). For the remainder of this paper we will be discussing brightness temperature differences derived from the interferometer data alone.

AIA provides full-disk images of the Sun at a number of EUV and UV wavelengths. We extracted the region of the ALMA observation from these full-disk images for further analysis. The spatial resolution of the AIA images presented in this paper is 1.52 $\sim$ 1.74\arcsec\ \citep{AIACal12}. The time cadence of the EUV images is 12 seconds, and that of the UV continuum images (1700 \AA) is 24 seconds.

XRT on Hinode also observed the active region simultenously. In this paper, we use the XRT images observed with the Al-thin filter. The spatial resolution and time cadence of the XRT data are 2\arcsec \ and 30 seconds, respectively.

The co-alignment between the ALMA images and the other wavelengths is achieved by the conversion of the a priori coordinate system of the ALMA images from RA/DEC to heliocentric. While the precision of the co-alignment is better than 2\arcsec \ for ALMA with respect to AIA, the positional uncertainty between ALMA and XRT might be as poor as 5\arcsec \ because it requires visual inspection, and there are not always matching features in the two very different temperature ranges contributing at millimeter and X-ray wavelengths.

\section{Plasmoid ejection observed with ALMA, AIA, and XRT}\label{sec:plasmoid}

\begin{figure}
\epsscale{1.1}
\plotone{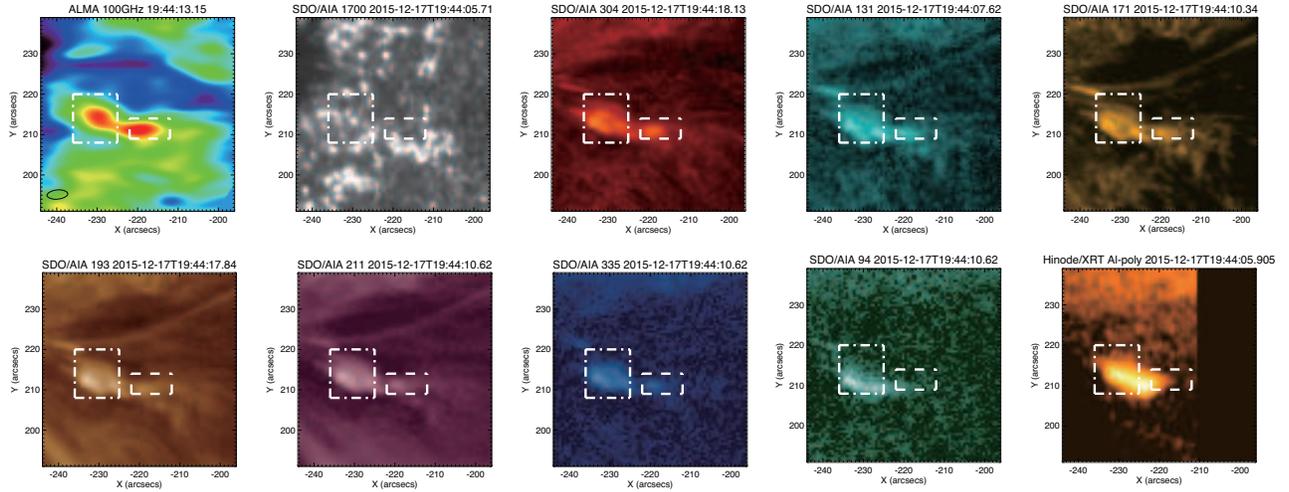}
\caption{Images of the plasmoid ejection at 19:44:13 UT, close to the intensity peak of the second ejection. The panels are labelled by the appropriate wavelength: in order from left to right and top to bottom, they are ALMA 100 GHz, AIA 1700, 304, 131, 171, 193, 211, 335 \& 94 \AA, and XRT soft X-rays with the Al-poly filter. The region shown corresponds to the dashed box in Figure \ref{fig:fig1}. The dash-dotted and dashed boxes in this figure are used to generate the light
curves shown in Figure \ref{fig:fig3}. \label{fig:fig2}}
\end{figure}

An X-ray bright point (XBP) is located at the leading penumbra of the large sunspot in NOAA12470, on the left side of the box marked in Figure \ref{fig:fig1}. Two tiny flares, not sufficiently bright to be classified as GOES soft X-ray events, occurred in the XBP (indicated by the dotted-dash box in Figure \ref{fig:fig2}) during the ALMA observation on 17 December 2015, at 19:39:30UT and 19:41:40UT. At the spatial resolution of AIA, Figure \ref{fig:fig2} shows that the XBP is spatially resolved into several small loops with significant substructure.

\begin{figure}
\epsscale{0.7}
\plotone{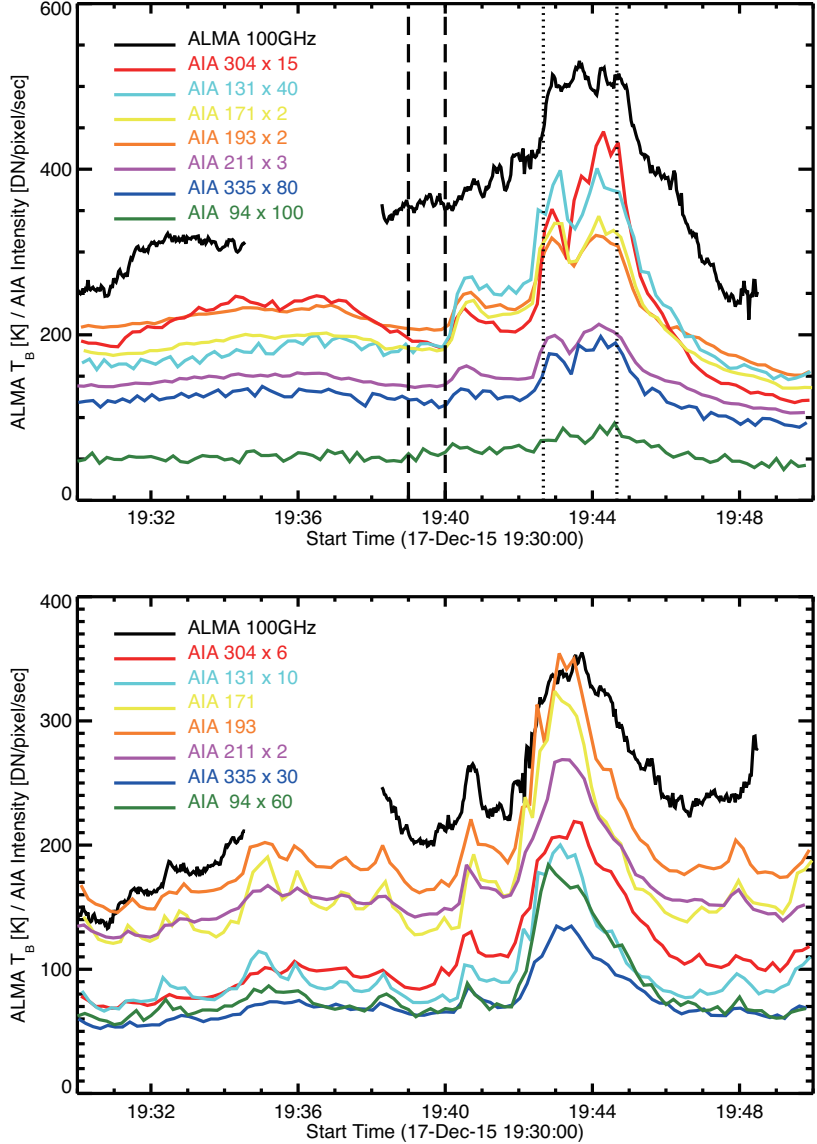}
\caption{Intensity profiles of the XBP and plasmoid in the AIA EUV bands and in the 100 GHz ALMA data. Upper panel: The total intensity integrated in the area indicated by the dashed box in Figure 2, corresponding to the motion of the plasmoid. Lower Panel: The total intensity of the (stationary) XBP in the area indicated by the dash-dotted box in Figure 2. There is no 100 GHz data between 19:34:33UT and 19:38:16UT when calibration scans were being run. \label{fig:fig3}}
\end{figure}

From Movie \ref{mov:mov1}, the tiny flare starting at 19:39:30 UT produces a thin jet that can be seen in all EUV bands. It travels away from the sunspot at an angle very slightly north of due west. The jet is readily seen in the 100 GHz images, as shown by the (red) contours in the middle-right region of the frame. The brightenings in the XBP during the movie occur at different spatial locations, and this ejection seems to be associated with an EUV brightening at the north-western edge of the XBP. The intensity-time profile of the XBP in the EUV bands and at 100 GHz (lower panel of Figure \ref{fig:fig3}) shows a small peak in intensity at about 19:40:30UT, while the western area encompassing the ejecta shows only a small increase when averaged over the area of the corresponding box. Since this event is weak, and the intensity of the jet structure is small, in the following discussion we focus on the second event.

The second event started at 19:41:40 UT with EUV brightening along the southern edge of the XBP, and we can see in Movie \ref{mov:mov1} that a plasmoid is again ejected from the XBP and travels away from the large sunspot, this time at an angle just south of due west. Both this plasmoid and the previous ejection appear to be moving at a large angle to the local vertical, although we have no way of determining the exact path in three dimensions. The size of the moving feature in the EUV images is 4\arcsec \ $\times$ 7\arcsec, i.e., similar to the spatial resolution of the ALMA images. The plasmoid moves at a projected velocity of about 40 km/s, and passes through the area indicated by the dashed box to the right of the XBP in Figure \ref{fig:fig2}. Figure \ref{fig:fig2} shows the plasmoid and XBP when the total intensity integrated over the western box for the plasmoid is near its peak (19:44:13 UT). The plasmoid is clearly visible in all the AIA EUV images with the exception of 94 \AA, which is quite noisy. It is also seen in the 100 GHz images, and in fact is brighter than the XBP at the time shown. On the other hand, the X-ray images obtained with XRT do not show any moving feature corresponding to the plasmoid, although there is a short stationary jet-like structure protruding to the west, matching the brightening in the XBP itself. When the plasmoid separates significantly from the XBP ($\sim$19:45:30UT), the XRT images do not show any corresponding signature above the noise level in the images. Based on the temperature response of XRT \citep{XRTCal11} we conclude that the plasmoid does not include significant coronal plasma at temperatures $>$1 MK. We also note that there is no enhancement in the AIA UV continuum image (1700 \AA) during the events. This confirms that the event is not predominantly chromospheric. 

\section{Thermal structure of the plasmoid}\label{sec:therm}

When we consider the thermal structure of the plasmoid, we have three basic working hypotheses to consider: 1) The plasmoid consists of (roughly) isothermal plasma that is optically thick at 100 GHz. 2) The plasmoid consists of isothermal plasma that is optically thin at 100 GHz. 3) The plasmoid consists of multi-thermal plasma. We examine which hypothesis is consistent with the intensity enhancements in the AIA and ALMA images. To proceed, we assume that the line-of-sight depth of the plasmoid is 4\arcsec \ ($\sim$3000 km), i.e., the same as the width of the plasmoid. The intensity enhancement corresponding to the plasmoid is derived from the time-averaged intensities during the period indicated by the dotted lines in the upper panel of Figure \ref{fig:fig3}, after subtracting a background level determined by averaging over the period shown by the dashed lines. We use only the data of the AIA bands where the averaged intensity is significantly larger than the dark-noise and read-noise level (the 171, 193, and 211 \AA \ bands). Although the 304 \AA \ data also satisfy this criterion, we cannot obtain reliable quantitative results from 304 \AA \ for the reasons given earlier. The averaged enhancements that should be explained by the hypotheses are 145 K, 68.1 DN/sec/pixel, 49.8 DN/sec/pixel, and 20.2 DN/sec/pixel at 100 GHz, 171, 192, 211 \AA \ AIA bands, respectively.

\begin{figure}
\epsscale{0.8}
\plotone{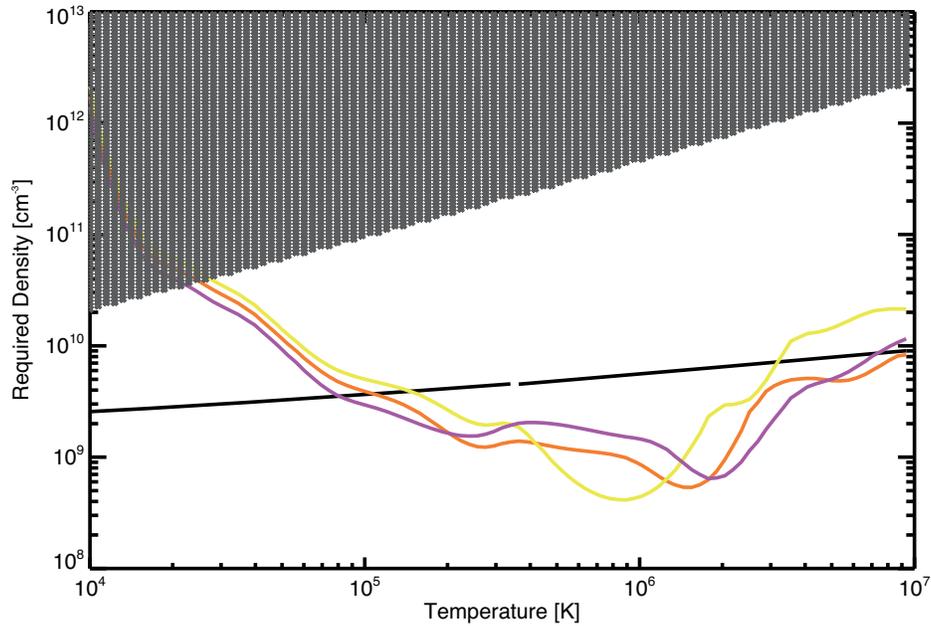}
\caption{Density required to explain each of the intensity enhancements of the plasmoid. The shaded region in the upper portion of the figure indicates conditions under which 100 GHz emission becomes optically thick. The black solid line indicates the density required for the observed brightness temperature enhancement at 100 GHz when we assume that the plasmoid is optically thin at the corresponding temperature. The gap in the black line is at the temperature where the Gaunt factor in the formulas changes (3.6$\times$10$^{5}$ K). Yellow, orange and purple lines indicate the required densities for the intensity enhancements in the 171, 193 and 211 \AA \ AIA bands, respectively.\label{fig:fig4}}
\end{figure}

\begin{figure}
\includegraphics[angle=270,width=\textwidth]{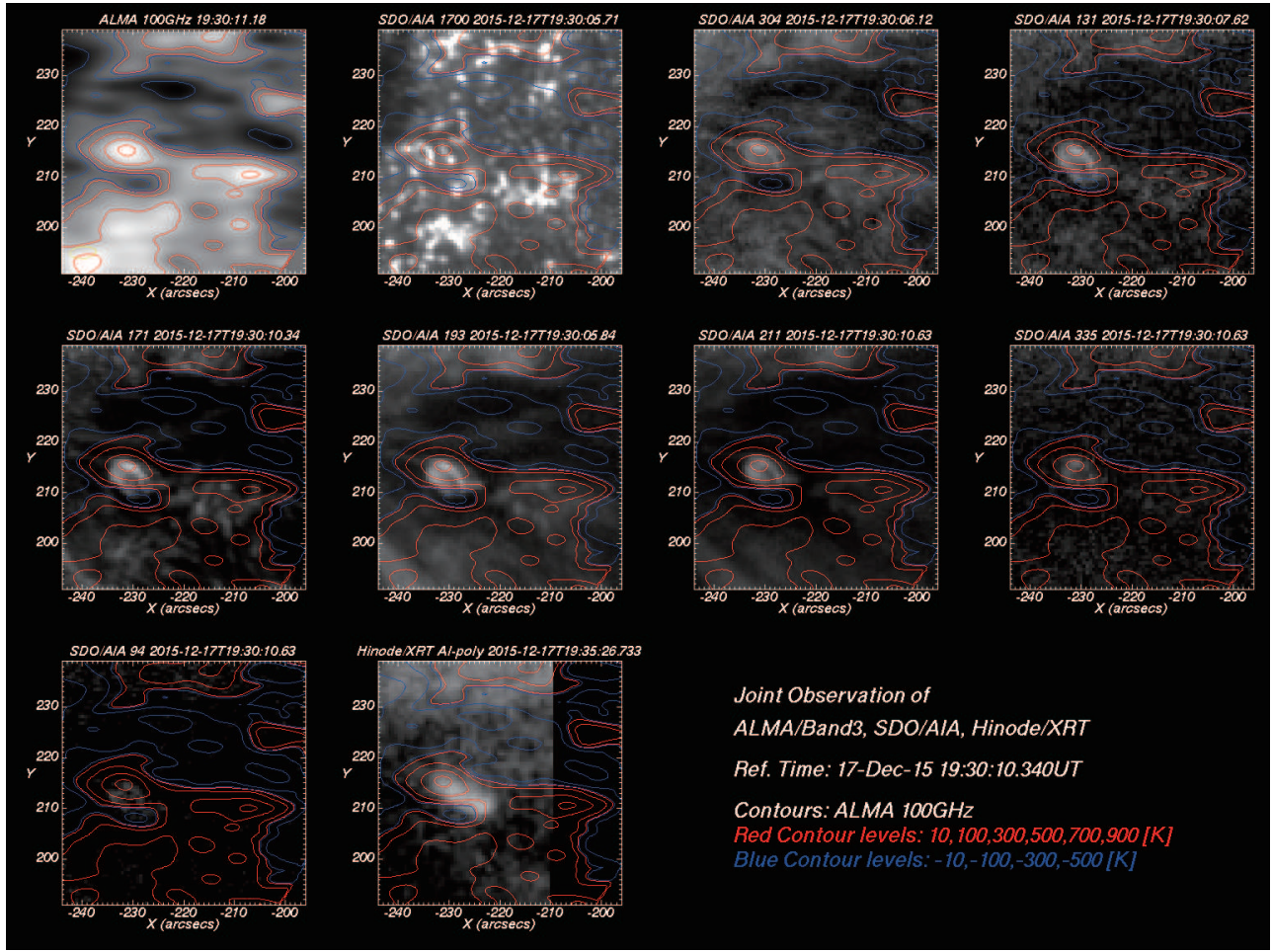}
\caption{Example image from the movie of the plasmoid ejection observed with ALMA/Band3, SDO/AIA, and Hinode/XRT. The contours indicate the relative brightness temperature in the 100GHz image; red and blue show positive and negative values, respectively. The movie file is available in the electronic edition.\label{mov:mov1}}
\end{figure}

Figure \ref{fig:fig4} shows the density required to explain the observed intensity enhancement at each wavelength when we assume that the plasmoid consists of isothermal plasma. The black solid line indicates the density required for the enhancement of 100 GHz when we assume that the plasmoid is optically thin at the corresponding temperature. To calculate the optical depth for 100 GHz, we use the standard formula for thermal free-free emission \citep[e.g.,][]{Lang80}, which is proportional to $n_{e}^{2}T^{-0.5}\lambda^{2}$ where $n_{e}$ is the ambient electron density, $T$ the temperature, and $\lambda$ the wavelength. This expression is multiplied by the usual weakly temperature- and wavelength-dependent approximation to the Gaunt factor. Yellow, orange and purple lines indicate the required densities for the intensity enhancements of the 171, 193 and 211 \AA \ bands, respectively. The plotted values are estimated from the temperature responses of each band \citep{AIACal12}. For the calculation, we assume that the filling factor of the plasma is unity. At high densities, as indicated by the shading in the upper region of the figure, the plasmoid becomes optically thick at 100 GHz. Based on this figure, we can now address the three hypotheses described above:

\paragraph{Hypothesis 1): Isothermal plasma that is optically thick at 100 GHz}
If the plasma is optically thick, the electron temperature of the plasmoid must be of order 10$^{4}$ K based on the temperatures in the combined image in Figure \ref{fig:fig1}. This regime lies at the left edge of Figure \ref{fig:fig4}, requiring a density larger than 2$\times$10$^{10}$ cm$^{-3}$. Based on Figure \ref{fig:fig4}, at this temperature densities of order 10$^{12}$ cm$^{-3}$ would be required to explain the AIA measurements. However, such a density would be larger than values typically found in dark filaments and would likely be observed as dark absorbtion features in the EUV images \citep{Hein08}. Since the plasmoid is observed to be in emission at all AIA EUV wavelengths, it cannot be an optically-thick 10$^{4}$ K isothermal source at 100 GHz.

\paragraph{Hypothesis 2): Isothermal plasma that is optically thin at 100 GHz}
In this case, ideally the temperature and density of the plasmoid would be indicated by a region where the black, yellow, orange and purple lines all cross. There is no point at which all four lines cross exactly, but they do gather together in the region around a temperature of 8.0$\times10^{4}\sim$1.4$\times$10$^{5}$ K and density of 3.5$\sim$3.9$\times$10$^{9}$ cm$^{-3}$. The temperature range includes the peak formation temperature of the He II 304 \AA \ line, and is far too low to contribute significantly at soft X-rays, thus being consistent with the absence of a clear signature of the plasmoid in the XRT images. Therefore, the hypothesis is reasonable, and an isothermal plasma with temperature 10$^{5}$ K and density 4$\times$10$^{9}$ cm$^{-3}$ could explain the intensity enhancements in both the AIA and 100 GHz images.

\paragraph{Hypothesis 3): Multi-thermal plasma}
We argued above that the 100 GHz emission cannot be from an optically thick isothermal source because such a dense and cool plasma would appear as a dark absorption feature at EUV wavelengths. If we drop the requirement for isothermality, then we can envisage a scenario in which the 100 GHz emission comes from a cool dense core, while the EUV emission arises in a hot sheath surrounding the core. In the plasmoid picture, the outer magnetic fields of a magnetic island that wrap the cool core formed by magnetic reconnection would be a natural location for such a sheath \citep[e.g.][]{Kaneko15}. The hot sheath needs to contribute only to the enhancements at the EUV bands. Accepting the argument based on XRT, the EUV-emitting plasma has to be cooler than 1 MK. Moreover, the plasma has to be hotter than a few$\times$10$^{5}$ K, because the 100GHz emission from $\le$10$^{5}$ K plasma that can explain the enhancements at the AIA bands is larger than the observed one. Hence, we argue that a temperature in the range 5$\times$10$^{5}$ K $\sim$ 1 MK would be a good match for the hot sheath. Since the radio emission is proportional to the square of the density, in comparing with the 100 GHz emission from the cool core, the contribution to 100 GHz emission from the hot sheath is negligibly small ($\le$10 K) when we assume that the density of hot sheath is 10$^{9}$ cm$^{-3}$. In both cases that the 10$^{4}$ K plasma forming the the cool core is either optically thin or thick at 100 GHz, the observed enhancement at 100 GHz can be explained. The density of the cool core would be $\ge$2$\times$10$^{10}$ cm$^{-3}$ for the optically thick case, or $\sim$10$^{9}$ cm$^{-3}$ for the optically thin case. The He II line can be formed over a wide temperature range (1$\times$10$^{4}\sim$5$\times$10$^{5}$ K) and thus the detection of the plasmoid in the 304 \AA \ AIA image is still consistent with this picture.

Based on the discussion above, the intensity enhancements of the plasmoid at 100 GHz and in the EUV bands of AIA/SDO (171, 193, and 211 \AA \ bands) can be explained as $\sim$10$^{5}$ K isothermal plasma that is optically thin at 100 GHz, or as a 10$^{4}$ K core with a hot envelope.

If we adopt the optically-thin parameters with the plasmoid at a temperature of 10$^{5}$ K, a volume of 4\arcsec$\times$4\arcsec$\times$7\arcsec= 3$\times$10$^{25}$ cm$^{3}$ and density of 3$\times$10$^{9}$ cm$^{-3}$, the mass of the plasmoid is of order 2$\times$10$^{11}$ g and the thermal and kinetic energies are 5$\times$10$^{24}$ ergs and 2$\times$10$^{24}$ ergs, respectively. These values are an order of magnitude smaller than, for example, the averaged thermal energy of polar X-ray jets reported by \citet{Sako13}, and much smaller than the recurrent active-region jets (masses $\ge$10$^{14}$ g) reported by \citet{Liu16}, indicating that this active-region event was a relatively small jet and demonstrating the sensitivity of ALMA for the study of such small energy releases.

This analysis demonstrates the value of the additional temperature and density constraints that ALMA can provide. In addition, at higher frequencies the spatial resolution of ALMA is close to that achieved with space-borne telescopes such as AIA (and significantly better than the XRT resolution), providing the ability to spatially resolve the same features at very different wavelengths providing complementary physical information. The ALMA science observations in Cycle 4 will have the improved spatial resolution \citep{Shimojo17}, and those data will then offer, e.g., the ability to distinguish between the two models we have arrived at here, using a detailed comparison of the millimeter-wavelength and EUV morphologies: if the shape of the 100 GHz source is the exactly same as that of the EUV source, that would favor the isothermal model, whereas an EUV source size larger than the ALMA measurement would favor the presence of a hot sheath. In addition, observations of a plasmoid at two frequencies simultaneously will be able to distinguish optically thick and thin contributions from their radio spectra. This is one example of the areas of study opened up by the availability of ALMA observations of the Sun.

\acknowledgments
This paper makes use of the following ALMA data: ADS/JAO.ALMA\#2011.0.00020.SV. ALMA is a partnership of ESO (representing its member states), NSF (USA) and NINS (Japan), together with NRC (Canada) and NSC and ASIAA (Taiwan), and KASI (Republic of Korea), in cooperation with the Republic of Chile. The Joint ALMA Observatory is operated by ESO, AUI/NRAO and NAOJ. SDO is the first mission to be launched for NASA's Living With a Star (LWS) Program. Hinode is a Japanese mission developed and launched by ISAS/JAXA, with NAOJ as domestic partner and NASA and STFC (UK) as international partners. It is operated by these agencies in co-operation with ESA and NSC (Norway).  M.S. was supported by JSPS KAKENHI Grant Number JP17K05397. This work was carried out on the solar data analysis system and common-use data-analysis computer system operated by ADC/NAOJ.

\end{document}